\documentclass[twoside,12pt]{article}

\usepackage{epsfig,rotating}

\def\Journal#1#2#3#4{{#1} {#2} (#4) #3 }

\def\NPA{{\em Nucl. Phys.} A}

\def\PLB{{\em Phys. Lett.} B}

\def\PRL{\em Phys. Rev. Lett.}

\def\PRC{{\em Phys. Rev.} C}

\newcommand{\be}{\begin{equation}}

\newcommand{\ee}{\end{equation}}

\newcommand{\bea}{\begin{eqnarray}}

\newcommand{\eea}{\end{eqnarray}}

\begin{document}
\title{Medium Modifications of Mesons in Elementary Reactions and Heavy-Ion Collisions} 
\author{Volker Metag\\
II. Physikalisches Institut\\
University of Giessen, Heinrich-Buff-Ring 16\\
D-35392 Giessen, Germany}
\maketitle

\begin{abstract}
  {Experimental searches for modifications of vector mesons in the
  nuclear medium are reviewed. Data on $\rho,\omega$ and $\Phi$ mesons are
  presented. The results have been obtained in elementary reactions with
  proton and  photon beams as well as in heavy-ion collisions. Compared to the 
  free particle properties, the $\omega$ and $\Phi$ meson are found to drop
  in mass at normal nuclear matter density  by 9-14$\%$ and 3.5$\%$ whereas
  their widths are reported to increase by factors of about 16 and 3.6,
  respectively. For the $\rho$ meson, conflicting results on in-medium mass 
  shifts and broadening have been published. The experimental data are
  compared to recent model calculations.} 
 \end{abstract}
\hspace*{1cm}{\em Keywords:} meson production, in-medium modifications
\section{Introduction}

Widespread experimental searches for changes of hadron properties in a
nuclear environment were motivated by theoretical studies in the 80's and early
90's \cite{Pisarski,Meissner,Brown_Rho,Hatsuda_Lee}, 
predicting a close connection between in-medium modifications 
and chiral symmetry
restoration in hot and/or dense matter. Subsequent investigations revealed that 
the link between nuclear properties and QCD symmetries was not as direct as
originally envisaged. A connection between hadronic spectral 
functions and QCD condensates is, however, provided by QCD sum rules which
relate the integral over hadronic spectral functions to $<\bar q q>$ and
higher order condensates. Changes of condensates with temperature and
density associated with a partial restoration of chiral symmetry 
only constrain corresponding in-medium modifications of
hadronic spectral functions \cite{Leupold_1,Leupold_2,Leupold_3,Leupold_4}; 
hadronic models are still needed for specific predictions of hadronic 
properties in the medium. 

\section{Theoretical predictions}

\begin{figure}[h]
\centering{
\includegraphics[width=0.35\textwidth]{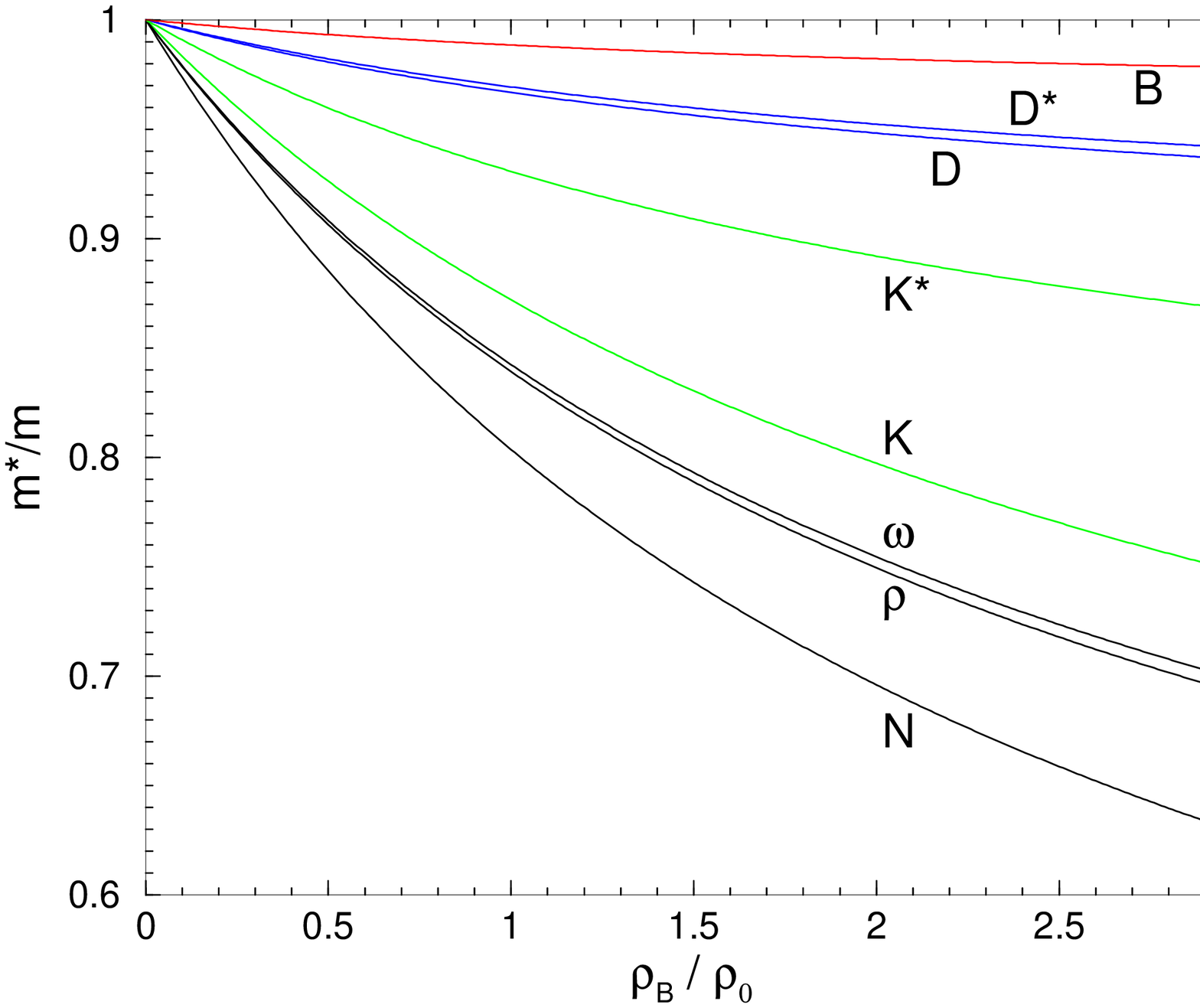}\hspace*{1.cm}
\includegraphics[width=0.50\textwidth]{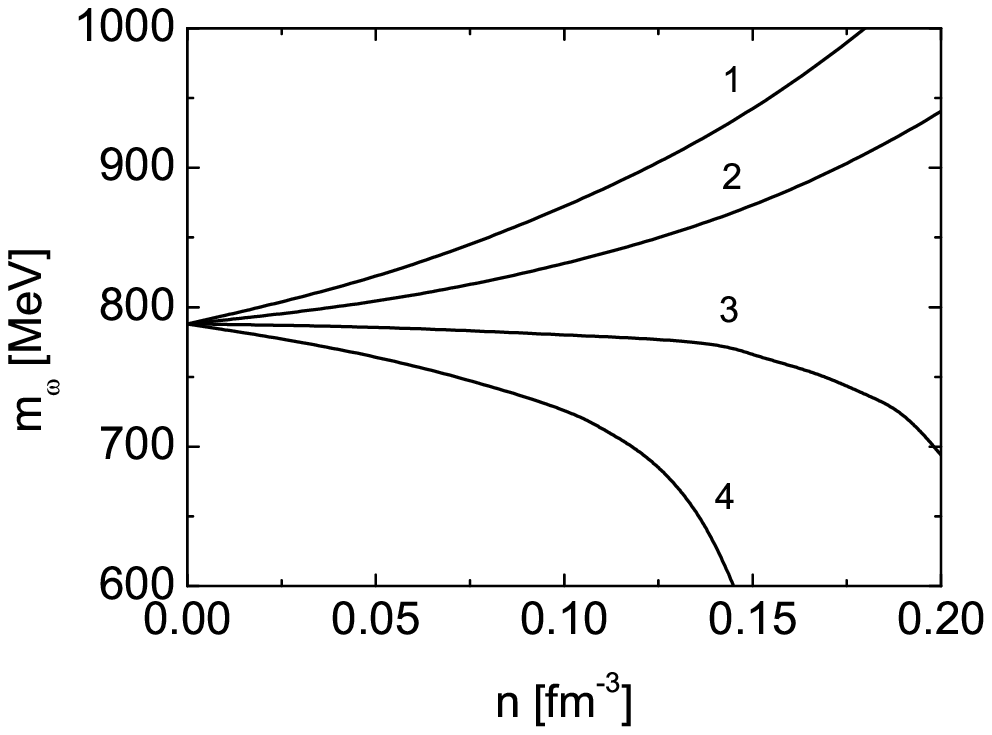}
\caption{Left: Hadron masses as a function of baryon density predicted within
  the QMC model \cite{QMC}. Right: Variation in the $\omega$ mass as a result of
  different density dependences of a 4-quark condensate \cite{Zschocke}.}
 \label{fig:prediction_1}.
}
\end{figure}
\begin{figure}[h]
\centering{
\includegraphics[width=0.45\textwidth]{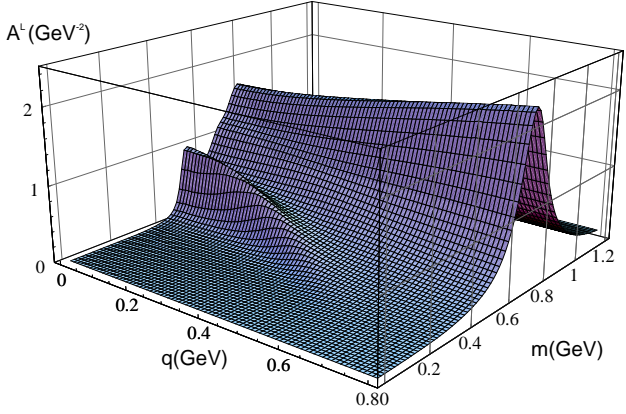}
\hspace{1cm}\includegraphics[width=0.41\textwidth]{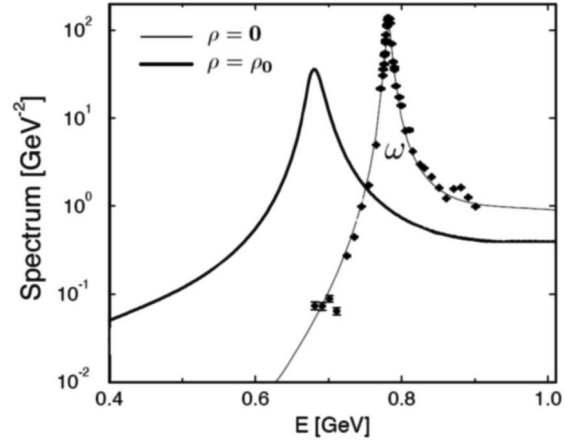}
\caption{Left: Modification of the $\rho$ spectral function with $\rho$
  momentum at normal nuclear matter density \cite{Peters}. 
 Right: The $\omega$ spectral function in vacuum and
 at normal nuclear matter density \cite{Klingl}.}
\label{fig:prediction_2}.
}
\end{figure}
\begin{figure}[h]
\centering{
\includegraphics[width=0.42\textwidth]{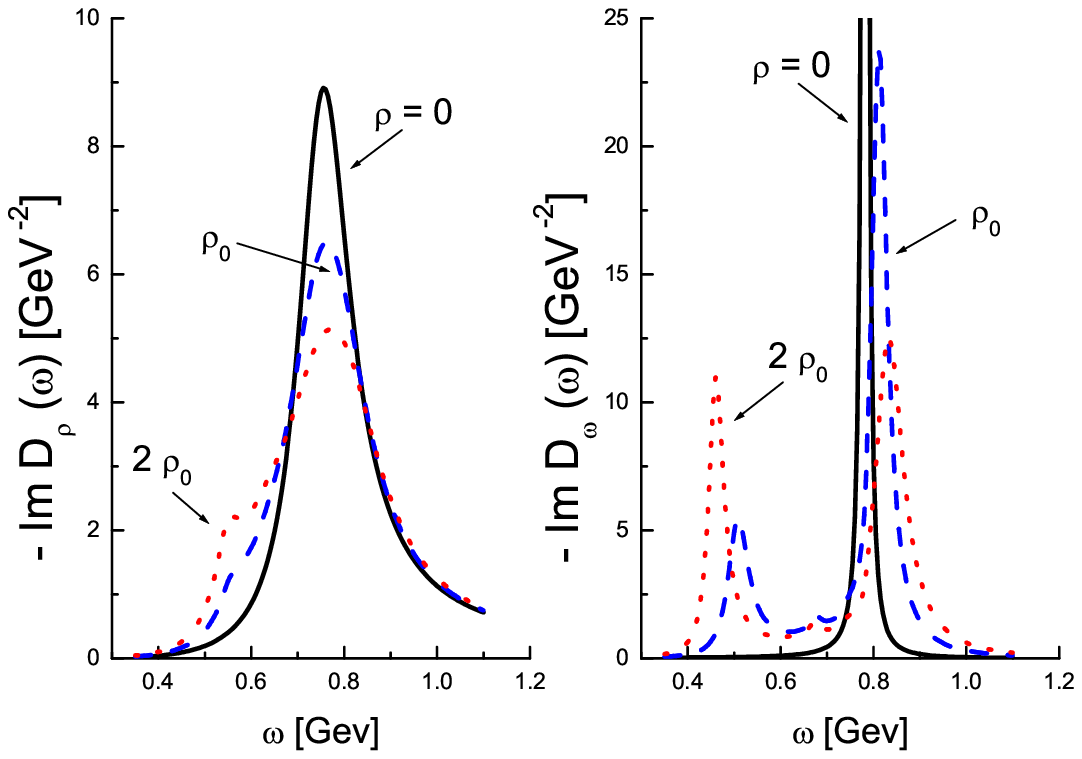}
\includegraphics[width=0.39\textwidth]{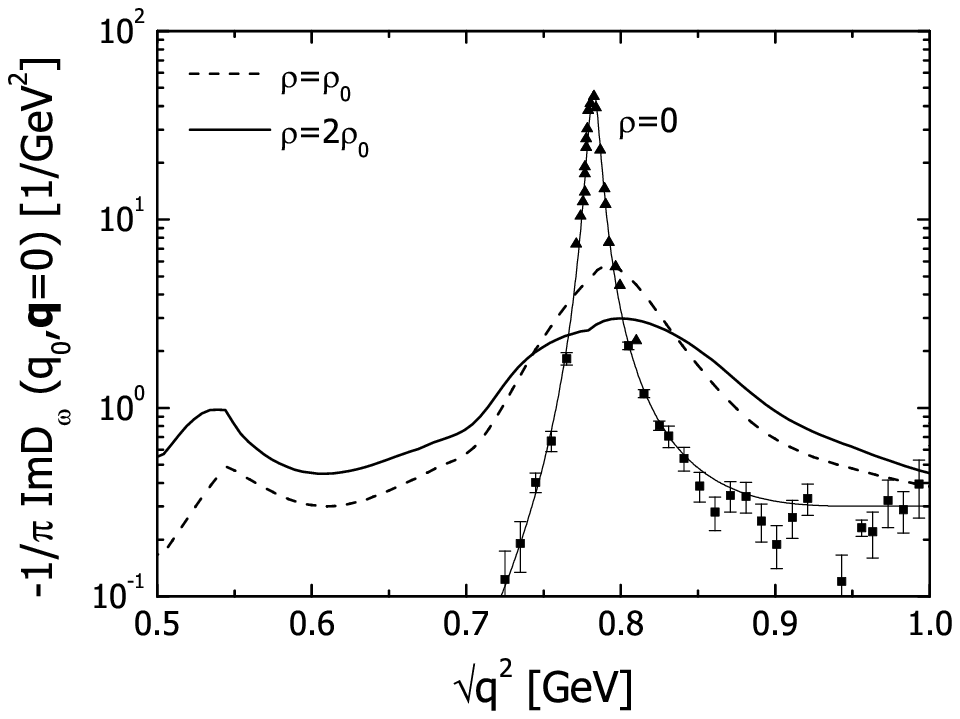}
\caption{Spectral functions at zero, normal, and twice normal nuclear matter
 density for the $\rho$ and $\omega$ meson \cite{Lutz} (left) and for the 
 $\omega$ meson \cite{Muehlich} (right). The structures at lower masses 
 arise from the coupling of the $\rho$ and $\omega$ meson to nucleon 
 resonances.}
\label{fig:prediction_3}.
}
\end{figure}

Many theory groups have developed different hadronic models to study the 
in-medium
behaviour of hadrons.  Some recent results of model calculations are
summarized in
Figs.~\ref{fig:prediction_1},~\ref{fig:prediction_2},~\ref{fig:prediction_3}.

In the QMC model a lowering of the $\omega$ mass by about 15$\%$ is expected
at normal nuclear matter density \cite{QMC}. The sensitivity of the $\omega$ 
mass to the
density dependence of some 4-quark condensate has been studied by Zschocke et
al. \cite{Zschocke}. Not only mass centroids but mass distributions have been
calculated: Klingl et al. \cite{Klingl} find a pronounced shift and broadening 
of the
$\omega$ mass in the nuclear medium. Structures in spectral functions arising
from the coupling of the respective meson to baryon resonances have been
predicted for $\rho$ and $\omega$ mesons by Lutz et al. \cite{Lutz}, Peters et
al. \cite{Peters} and M\"uhlich and Mosel \cite{Muehlich}. It should be noted that
- as shown in
Fig.~\ref{fig:prediction_2} - these structures fade out for meson momenta 
above several 100 MeV/c with respect to the nuclear medium. In order to be 
sensitive
to such structures in the experiment detector systems with sufficient
acceptance for low momentum mesons have to be provided. 

The variety of theoretical predictions calls for an experimental
clarification. Only recently, experiments have advanced to an accuracy
which allows to distinguish between different model predictions. Corresponding
experiments are described in the subsequent sections.

\section{Experimental approaches}

Experimentally, the predicted in-medium phenomena can be studied by measuring
the mass distribution of hadrons which are so shortlived that they decay
within the nuclear environment, i.e. in the atomic nucleus or in the collision 
zone of a heavy-ion reaction, after being produced in some nuclear reaction.
Information on the in-medium mass $m$ of hadron H can be deduced from the 4-momentum
vectors $p_1, p_2$ of the decay products X$_1$, X$_2$ for different 3-momenta $\vec{p}$ 
of the hadron with respect the nuclear medium. In general, the mass depends 
on the baryon density $\rho$ and temperature T of the medium:

\be
m(\vec{p},\rho,T) = \sqrt{(p_1 + p_2)^2}
\ee

Lepton pairs are the preferred decay channel because they escape even a
compressed collision zone of a heavy-ion reaction without strong final state 
interactions which would otherwise distort the 4-momentum vectors and thus the
determination of the invariant mass. The disadvantage of this decay mode is
the very small branching ratio of the order of $10^{-5} - 10^{-4}$ which makes
these measurements very difficult and sensitive to the background subtraction.

As pointed out by U. Mosel et al. \cite{Eichstaedt}, the experimentally determined 
invariant mass distribution (eq. 1) does, however, not directly provide the 
spectral function of the meson but rather represents a convolution of 
this spectral function $A(m)$ with the partial decay width $\Gamma_{H \rightarrow X_1,X_2}(m)$
into the channel being studied:

\be
\frac{d \sigma_{H \rightarrow X_1,X_2}}{dm} \sim A(m) \cdot \frac{\Gamma_{H
      \rightarrow X_1,X_2}(m)}{\Gamma_{tot}(m)}
\ee

Since $\Gamma_{H \rightarrow X_1,X_2}(m)$  depends itself on the invariant 
mass $m$ this may lead to deviations of the experimentally determined mass 
distribution from the true spectral function, in particular 
for broad resonance states.

Medium modifications have been investigated experimentally both in elementary
reactions and heavy-ion collisions. Both approaches have advantages and
disadvantages. Any signal from heavy-ion reactions represents an integration
over the full space-time evolution of the collision, involving strong
variations in densities and temperatures. On the other hand, a regeneration of
mesons in the collision zone helps to enhance the in-medium effects. In
elementary reactions on nuclei there is no time dependence of the density and
temperature which makes the theoretical analysis of the results much easier.
Because of the lower densities probed in these reactions medium effects may,
however, be less pronounced.

\subsection{Medium modifications studied in elementary reactions}
\begin{figure}[h]
\centering{
\includegraphics[width=0.41\textwidth]{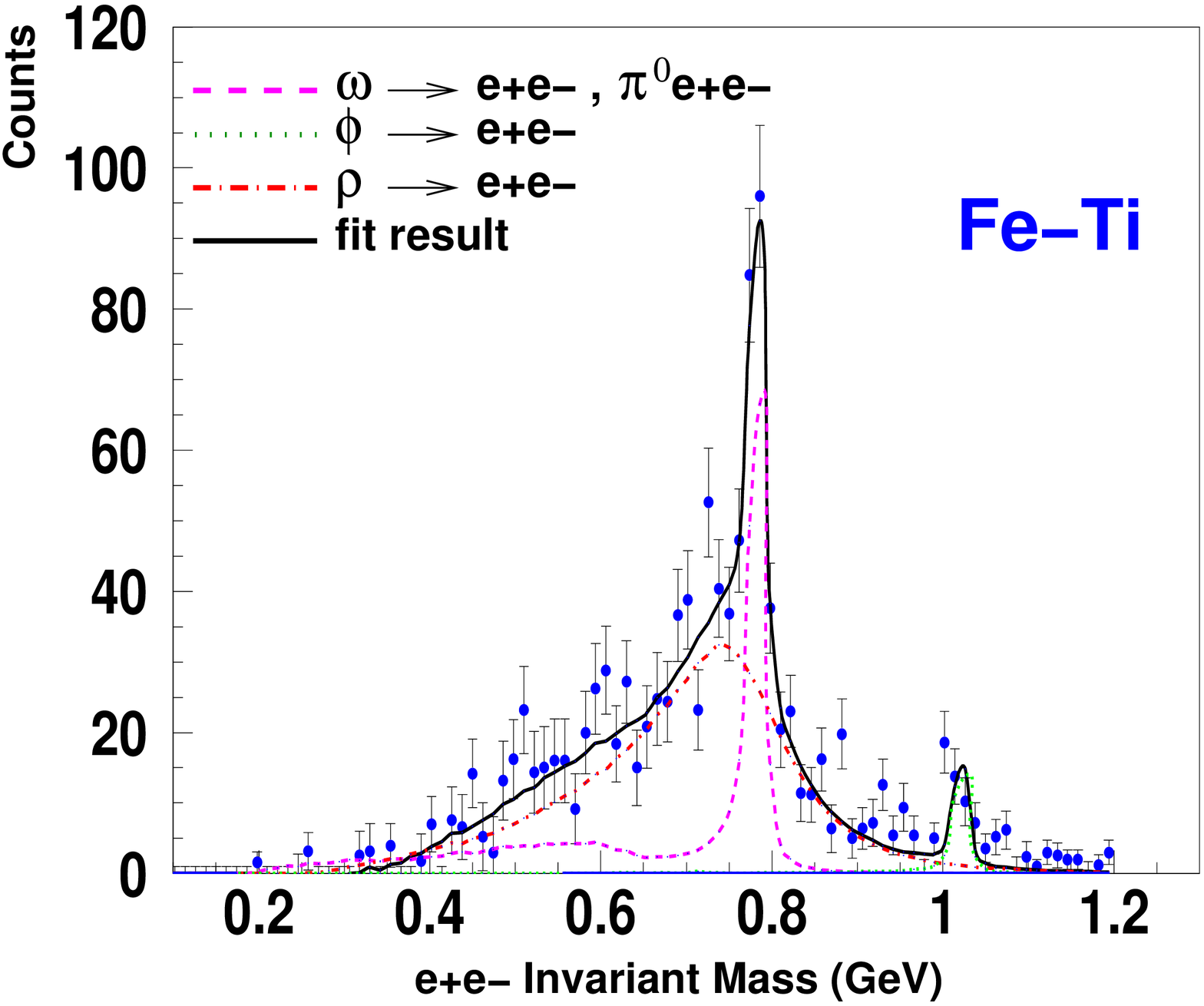}
\includegraphics[width=0.33\textwidth]{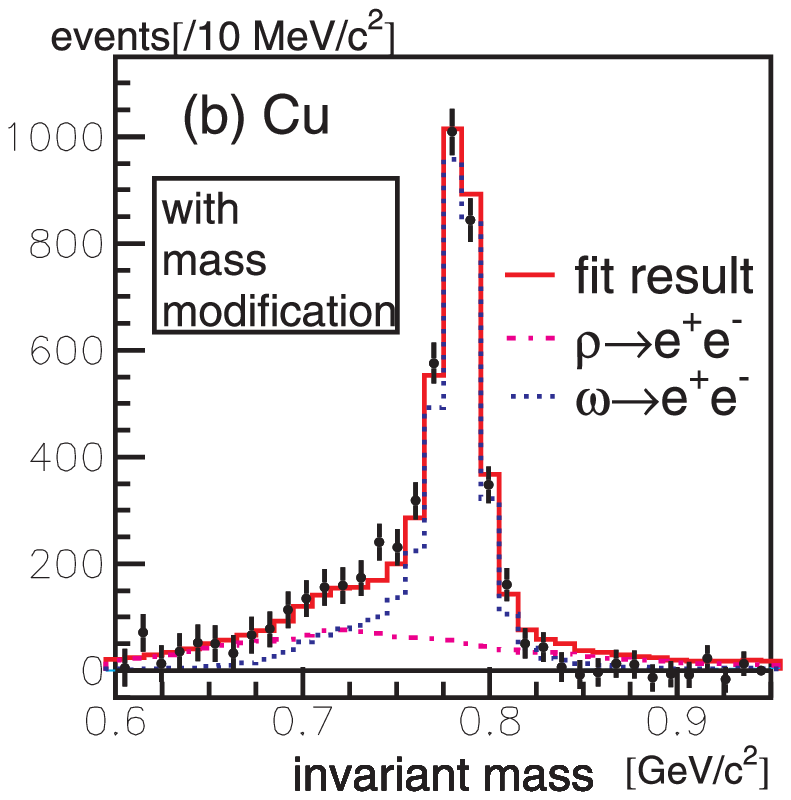}
\caption{$e^+e^-$ invariant mass spectra after background subtraction
    obtained (left) in photonuclear reactions ($E_{\gamma}$ = 0.6 - 3.8 GeV) 
    \cite{Djalali} and (right) in 12 GeV proton induced reactions 
    \cite{Naruki}. }
\label{fig:Jlab_KEK}.
}
\end{figure}

Proton as well as photon induced reactions have been studied to search
for medium modifications of mesons in elementary processes.
In-medium properties of the $\rho$ meson have been deduced from two
experiments at Jlab \cite{Djalali} and KEK
\cite{Naruki}, irradiating various targets with photon beams of 
E$_{\gamma}$ = 0.6 - 3.8 GeV and 12 GeV proton beams, respectively. 
The e$^+$e$^-$ invariant mass spectra resulting after subtraction of the 
combinatorial background are shown in Fig.~\ref{fig:Jlab_KEK}.
Despite of the similarity of both spectra both groups come to
conflicting conclusions: while the Jlab experiment \cite{Djalali} reports no
mass shift and a small in-medium broadening Naruki et al. \cite{Naruki} claim
a drop of the $\rho $ meson mass by 9$\%$ and no in-medium broadening.  
This discrepancy is most likely due to differences in treating the
combinatorial background and to details of the fitting procedures. In
particular, it appears important to normalize the combinatorial
background from mixed events at low invariant e$^+$e$^-$ masses
to that determined on an absolute scale from like sign pairs. 
It should be noted that both experiments have acceptance only for mesons with
3-momenta above 0.5-0.8 GeV/c and are thus not sensitive to possible 
in-medium
modifications as predicted in Fig.~\ref{fig:prediction_3} which are expected at
momenta less than 500 MeV/c.

\begin{figure}[h]
\centering{
\includegraphics[width=0.6\textwidth]{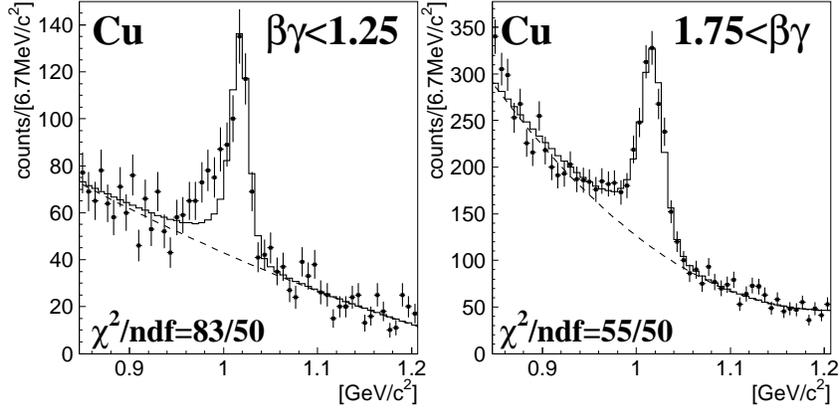}
\caption{e$^+$e$^-$ invariant mass spectra near the $\Phi$ mass peak for
 slow ($\beta \cdot \gamma < 1.25$) (left) and  fast $\Phi$ mesons
($\beta \cdot \gamma > 1.25$)(right) \cite{Muto}.}
\label{fig:KEK_Phi}.
}
\end{figure}

The KEK experiment has also investigated possible in-medium modifications of
the $\Phi$ meson (see Fig.~\ref{fig:KEK_Phi}) \cite{Muto}. A significant
excess on the low-mass side of the $\Phi$ meson peak is observed for slow 
$\Phi$ mesons ($\beta \cdot \gamma < 1.25$) which have a higher 
probability to decay in the Cu nucleus than fast $\Phi$ mesons. 
From an analysis of the structure in the sprectrum Muto et al. 
\cite{Muto}
extract a drop of the $\Phi$ mass by 3.4$\%$ and an increase of the $\Phi$
width by a factor 3.6 at normal nuclear matter density $\rho_0$.

The $\omega$ meson in the medium has been studied at much lower momenta 
($<$ 500 MeV/c) in a photoproduction experiment \cite{Trnka} at ELSA. Here,
the decay mode $\omega \rightarrow \pi^0 \gamma$ has been investigated which
has a much higher branching ratio of 9$\%$. Another advantage of this decay
mode is the insensitivity to possible in-medium
modifications of the $\rho$ meson ($\rho \rightarrow \pi^0 \gamma: 7 \cdot
10^{-4}$). A serious disadvantage, however, are possible strong
final state interactions of the $\pi^0$ meson within the nucleus which may
distort the extracted invariant mass distribution. Detailed Monte Carlo
simulations \cite{Messchendorp} show that this effect is small in the mass 
range of interest (600 MeV/c$^2 < m_{\pi^0 \gamma} < 800 MeV/c^2$)
and can even be further reduced by appropriate cuts.

The best way to identify possible in-medium modifications is to compare $\omega$
photoproduction on nuclei with a corresponding measurement on the proton which 
serves as 
reference. From the subgroup of 3 $\gamma$ events Trnka et al. \cite{Trnka}
deduced $\pi^0 \gamma$ invariant mass spectra. After fitting and subtracting 
the background the   
comparison of the corresponding invariant mass spectra for Nb and LH$_2$ targets 
exhibited a shoulder on the low mass side of the $\omega$ signal on the
nuclear target. This was taken as evidence for an $\omega$ in-medium mass
shift by 60$^{+10}_{-35}$ MeV at an average nuclear density of 
0.6 $\rho_0$. An extrapolation to normal
nuclear density leads to a drop in the $\omega$ mass by about 14$\%$.
\begin{figure}[h]
\centering{
\includegraphics[width=0.45\textwidth]{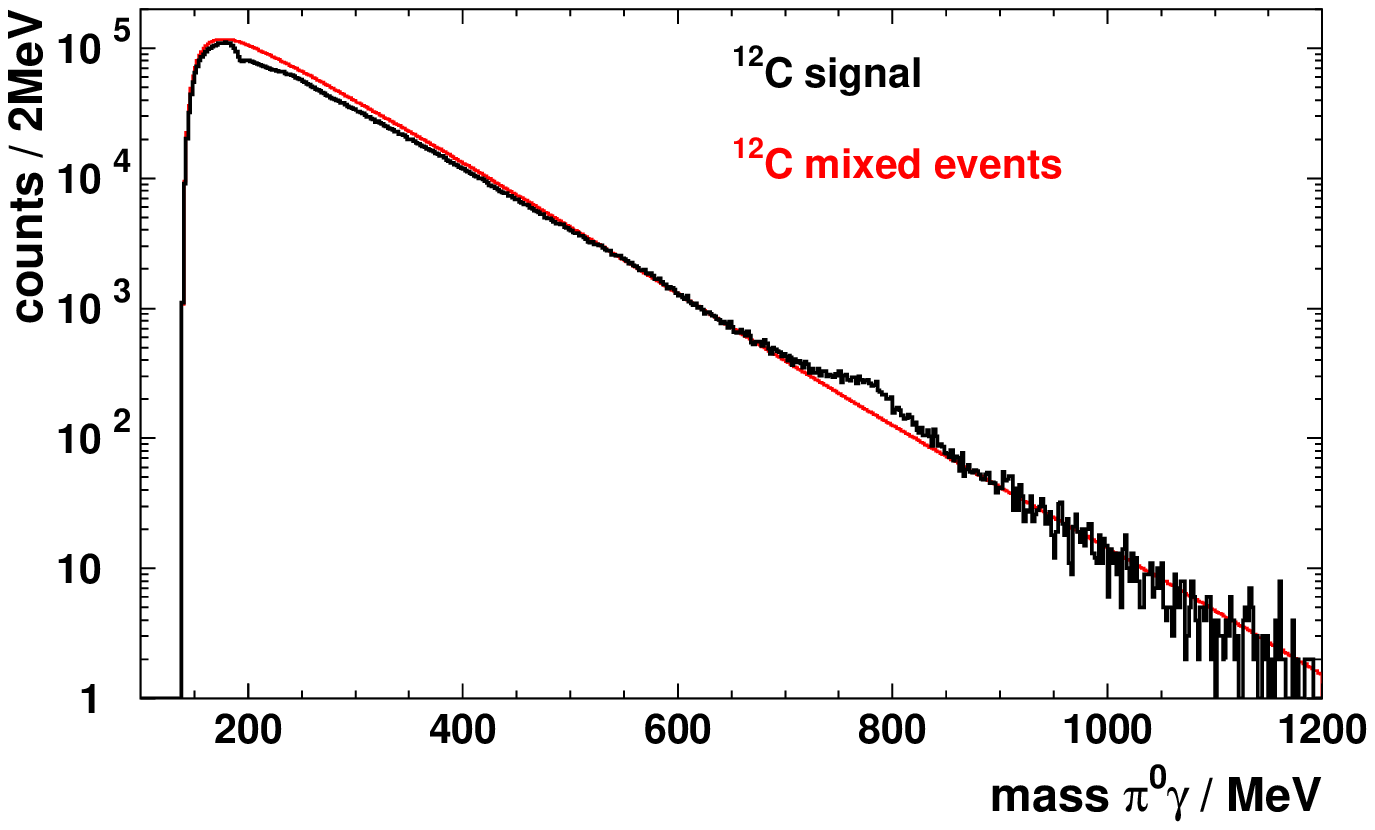}\hspace*{1.cm}
\includegraphics[width=0.45\textwidth]{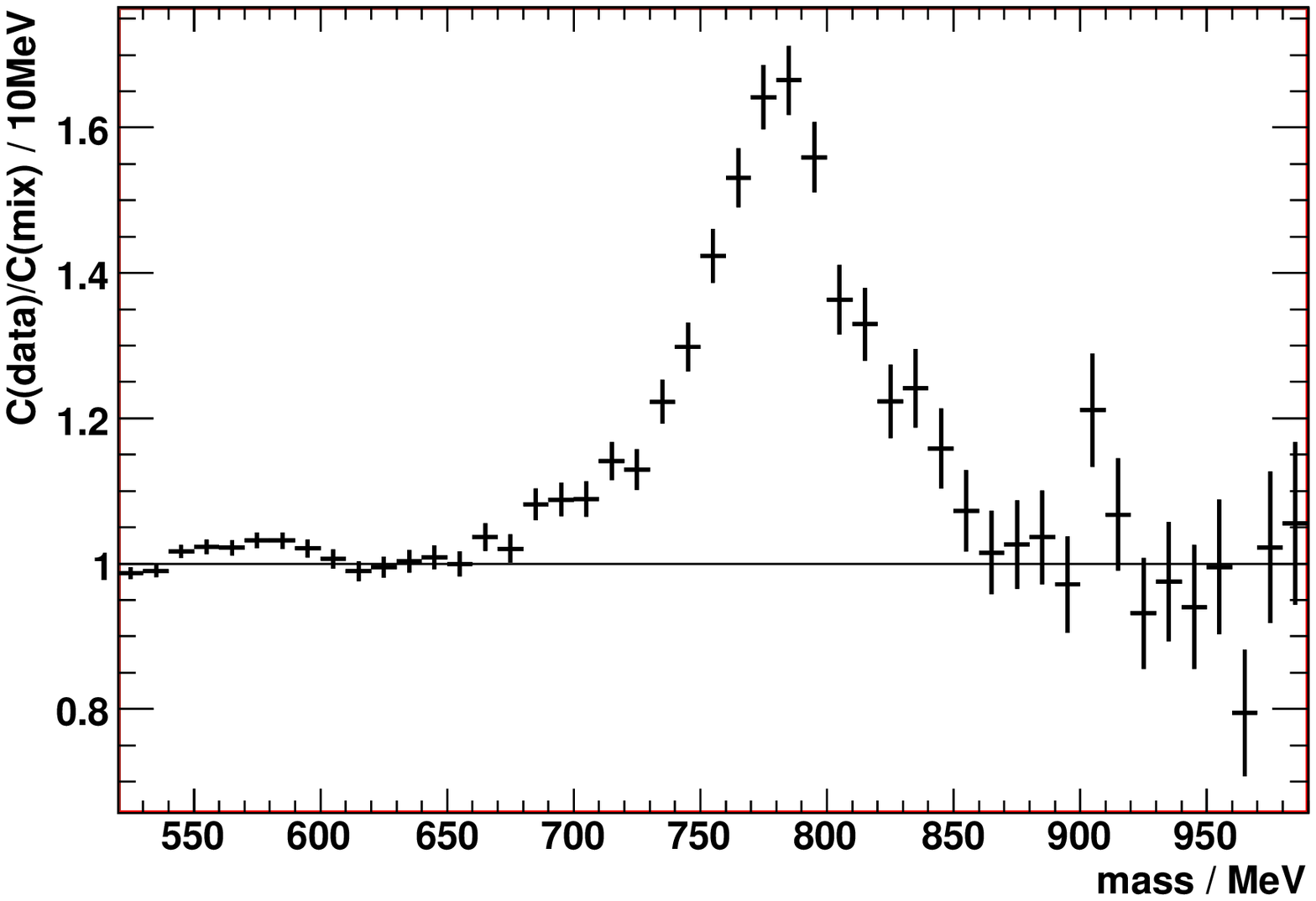}
\caption{Left: $\pi^0 \gamma$ invariant mass distribution for a photonuclear
 reaction (E$_{\gamma}$ = 0.7-2.5 GeV) on carbon in comparison to a mixed
 event background (see text) \cite{Kotulla_mix}. Right: ratio of the data to 
 the mixed event background.}
\label{fig:minv_mix_C}.
}
\end{figure} 

The shape of the $\omega$ signal is sensitive to the way the background is 
treated.
In \cite{Trnka} the background was fitted with an arbitrary function. In a
more rigorous treatment one could try to reproduce the background by summing up
all possible sources which can contribute to the $\pi^0 \gamma$ channel due to
limited acceptances and/or particle misidentification.
Another possibility is to determine the background with the mixed-event
technique as used in the lepton pair experiments. This approach has been
chosen for the analysis of new data taken on a carbon target. 
Fig.~\ref{fig:minv_mix_C}
shows the $\pi^0 \gamma$ invariant mass spectrum together with the
uncorrelated $\pi^0 \gamma$ background obtained by event-mixing 
\cite{Kotulla_mix}. Here, the 
invariant mass is calculated by combining a
$\pi^0$ from one event with a photon from another event. The mixed-event
background describes the experimental data over an invariant mass range of
about 400 MeV/$c^2$ with an accucracy of better than 5$\%$. This is
demonstrated in Fig.~\ref{fig:minv_mix_C} (right) which shows the ratio of 
the data to the mixed event background on a linear scale.
   
\begin{figure}[h]
\centering{
\includegraphics[width=0.35\textwidth]{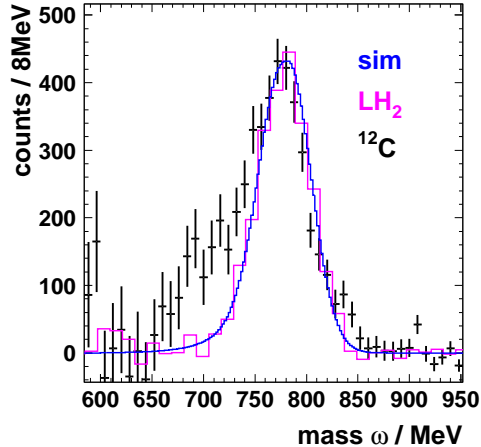}
\caption{$\pi^0 \gamma$ invariant mass distribution near the $\omega$ mass
  after subtracting the mixed event background of Fig.~\ref{fig:minv_mix_C}.
  For comparison the line shape from the corresponding measurement on the
  LH$_2$ target (histogram) and from a simulation (curve) are shown.}
\label{fig:minv-mix_C}.
}
\end{figure}

Subtracting this combinatorial background leads to the $\omega$ signal
shown in Fig.~\ref{fig:minv-mix_C} which again exhibits a shoulder on 
the low mass
side in comparison to the $\omega$ signal measured on the liquid
hydrogen target and to a simulation of the free $\omega$ signal, including the
experimental resolution. Thereby, the observation of an in-medium lowering of
the $\omega$ meson \cite{Trnka} is confirmed. To quantify the effect, the
$\omega$ signal has to be decomposed into an in-medium decay
and an in-vacuum decay contribution.
The lineshape of the in-vacuum decay component is known from the measurement 
on the LH$_2$ target, the in-medium
decay distribution is taken from BUU simulations \cite{Muehlich_priv}. 
A best fit to the signal is obtained by assuming a 
drop of the $\omega$ mass by 14 $\%$ at normal nuclear matter density in
accordance with \cite{Trnka}. 

Because of the sensitivity of the $\omega$ signal to the background treatment
an approach \cite{Oset,Kaskulov} to {\em assume} the background on the nuclear 
target to be the same as for the LH$_2$ target can only lead to wrong 
conclusions. The experimental data clearly show that the background 
distributions for the LH$_2$ and nuclear targets are different. This becomes
evident when one compares the background distributions over a wider mass range
than the limited one considered in \cite{Oset,Kaskulov}.

\begin{figure}
    \centering{
    \includegraphics[width=0.70\textwidth]{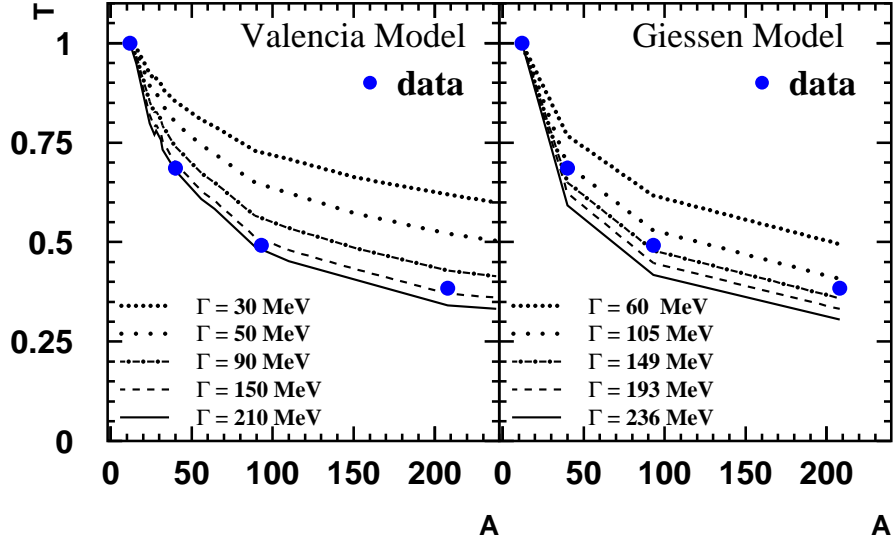}
    \caption{Transparency of nuclei for $\omega$ mesons. The preliminary data
    of the CBELSA/TAPS collaboration \cite{Kotulla} normalized
    to $^{12}$C are compared to calculations by  Kaskulov et
    al. \cite{Kaskulov} (left) and  M\"uhlich et al.
    \cite{MueMos}(right).}
    \label{fig:T_Gi_Val}
}
\end{figure}

Due to the detector resolution and uncertainties in the decomposition of the
$\omega$ signal in an in-medium and in-vacuum decay contribution it is
difficult to extract an in-medium $\omega$ width from the above experiment.
An access to the in-medium width of the $\omega$ is provided by measuring
the transparency ratio T 
\cite{Kaskulov,MueMos}
\begin{equation}
T=\frac{\sigma_{\gamma A \rightarrow \omega X}}{A \cdot \sigma_{\gamma N \rightarrow \omega X}},
\end{equation}
i.e. the ratio of the $\omega$ production cross section on a nucleus divided
by the number of nucleons A times the $\omega$ production cross section on a
free nucleon. As the $\omega$ photoproduction cross section on the neutron is
not yet known the transparency ratio is here normalized to carbon. If nuclei 
were completely transparent to $\omega$ mesons,
the tranparency ratio would be T=1. Consequently, T is a measure for the loss of
$\omega$ flux via inelastic processes in nuclei and can be determined in
attenuation experiments on nuclei of different mass A. Within the low density
approximation the $\omega$A absorption cross section $\sigma$ is related to the
inelastic $\omega$ width by $\Gamma_{\omega} = \hbar \rho v \sigma$. 
A comparison of preliminary data
from the CBELSA/TAPS collaboration \cite{Kotulla} 
with calculations of the Valencia \cite{Kaskulov} and Giessen
\cite{MueMos} theory groups 
(s. Fig~\ref{fig:T_Gi_Val}) 
yields an in-medium $\omega$ width in the nuclear reference frame of about 
130-150 MeV at normal nuclear matter density and at an average $ \omega$ 
momentum of 1100 MeV/c. This implies an in-medium broadening of the $\omega$ 
meson by a factor $\approx$ 16. Assuming the momentum dependence of the 
$\omega$ width given in \cite{MueMos} this value corresponds to a
total width of the $\omega$ meson at rest in the medium of about 70 MeV.

\subsection{Medium modifications studied in heavy-ion collisions}
The dropping $\rho$ mass scenario initially proposed in \cite{Pisarski,Brown_Rho,Hatsuda_Lee} was a prime motivation for the CERES collaboration to study
e$^+$e$^-$ pair emission in ultra-relativistic nucleus-nucleus collisions. 
\begin{figure}[h]
\centering{
\includegraphics[width=0.32\textwidth]{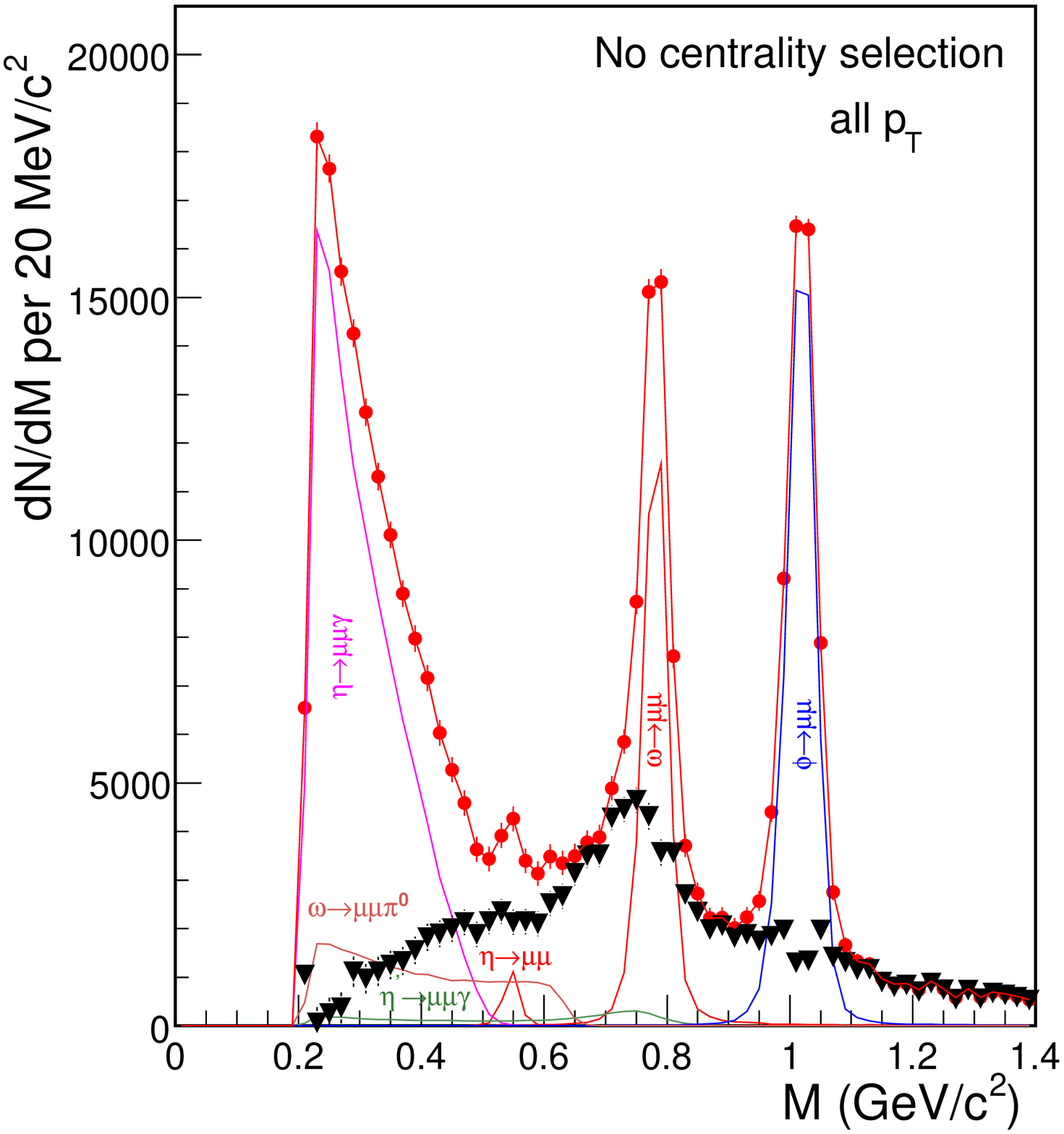}
\hspace*{0.1cm}
\includegraphics[width=0.29\textwidth]{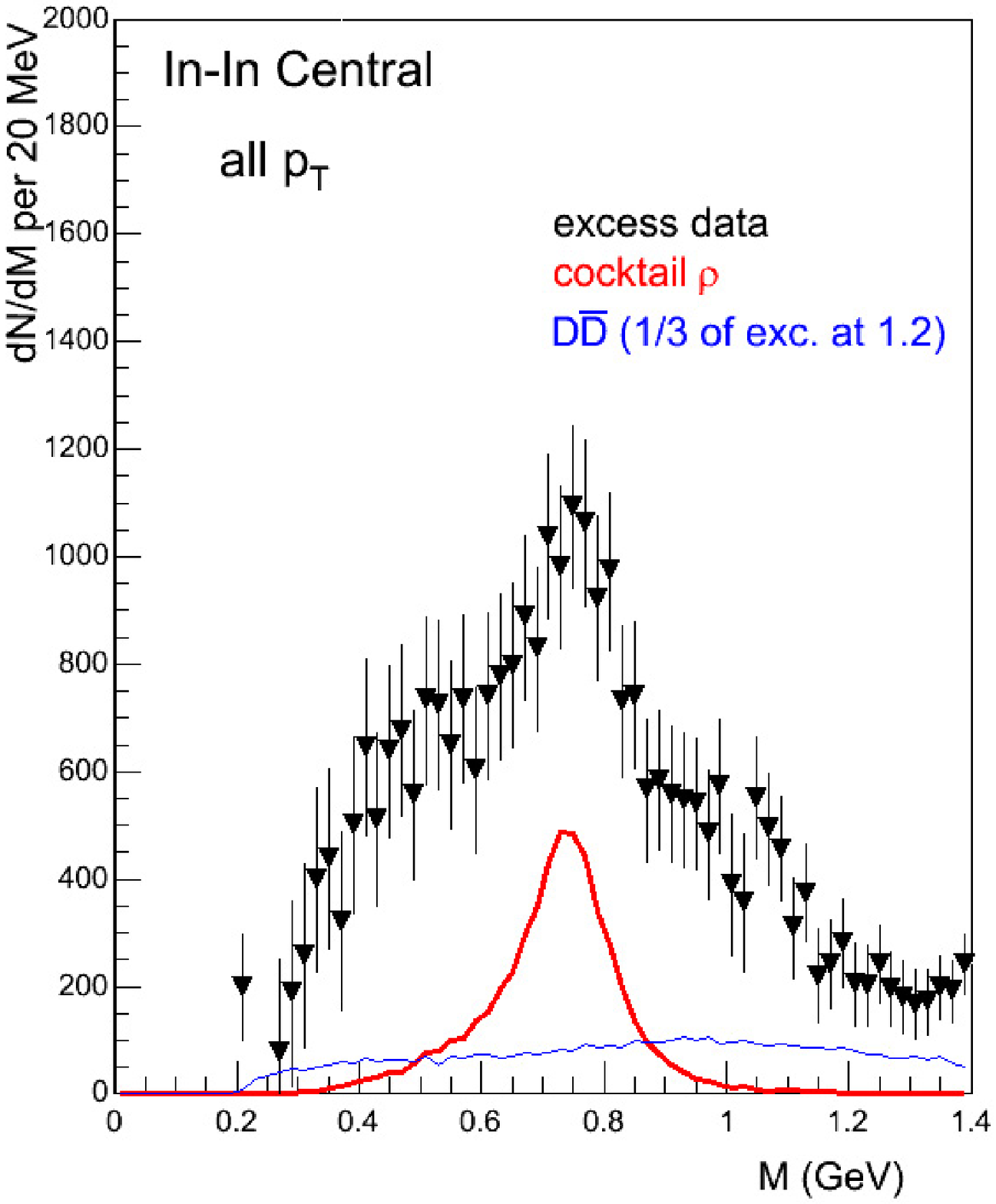}
\caption{Left: $\mu^+ \mu^-$ invariant mass spectrum from 158 AGeV In+In 
collisions after background subtraction. Right: Excess mass spectrum of
 dimuons for central In+In collisions in comparison to the spectral function
 of a free $\rho$ meson \cite{NA60}.}
\label{fig:NA60}.
}
\end{figure}
An excess in the dilepton yield above expectations from post-freeze-out 
leptonic decays of hadrons was observed \cite{CERES_95} and attributed to 
$\pi^+ \pi^- \rightarrow \rho \rightarrow e^+e^-$ annihilation via an
intermediate $\rho$. The dilepton excess is, however, smeared out over a 
much wider mass range than expected for the dropping $\rho$ mass scenario
\cite{Brown_Rho}. Adamova et al. \cite{CERES_06} come to the conclusion that
a broadening of the $\rho$ spectral function has to be favoured over a density
dependent downward shift in mass.

A breakthrough with regard to statistics and resolution in dilepton
spectroscopy of nucleus-nucleus collisions has been achieved by the NA60
collaboration \cite{NA60,Damjanovic} who studied the $\mu^+\mu^-$ decay channel in the
In+In reaction at 158 AGeV. Fig. ~\ref{fig:NA60} shows the $\mu^+\mu^-$
invariant mass spectrum after subtracting the combinatorial background. Peaks
from the $\omega$ and $\Phi$ decays are easily identified.
The quality of the data allowed to subtract the
{\em measured} post-freeze-out dilepton cocktail separately for different 
centrality bins. The
remaining di-muon invariant mass spectrum is attributed mainly to the
$\rho \rightarrow \mu^+\mu^-$ decay. Fig.~\ref{fig:NA60} (right) 
indicates a strong in-medium broadening but
no mass shift of the $\rho$ meson. S. Damjanovic et al. \cite{Damjanovic}
conclude that the dropping mass scenario \cite{Brown_Rho}
is incompatible with the experimental data. 

{\hspace*{-1.5cm}
\begin{table}
\caption{Compilation of experimental results on the in-medium mass and width
of the $\rho, \omega, $ and $\Phi$ meson, measured in different experiments.
The production reaction and the momentum acceptance of the respective detector
system are given.}

\vspace*{0.2cm}

{\hspace*{-0.8cm}
\begin{tabular}{|c|c|c|c|c|c|}

\hline
 & KEK & Jlab & CBELSA/TAPS & CERES & NA60\\
\hline
reaction& pA 12 GeV & $\gamma$A 0.6-3.8 GeV & $\gamma$A 0.7-2.5 GeV& Au+Au 158
 AGeV & In+In 158 AGeV\\
\hline
momentum & & & & & \\ 
acceptance &  $p>$ 0.5 GeV/c &  $p>$ 0.8 GeV/c& $p>$ 0.0 GeV/c & $p_t>$ 0.0 GeV/c& $p_t>$ 0.0 GeV/c \\ 
\hline
 & & $\Delta m \approx$ 0 & & broadening
 favoured & $\Delta m \approx$ 0\\
$\rho$ & $\frac{\Delta m}{m} = - 9\%$ & some  & & over density & strong \\
& no & broadening & & dependent mass shift & broadening\\
\cline{1-1} \cline{3-6}
$\omega$ & broadening & &$ \frac{\Delta m}{m} \approx - 14 \%$ & &   \\
    & &    &$ \frac{\Gamma_{\omega}(\rho_0)}{\Gamma_{\omega}} \approx $16 & & \\
\hline
$\Phi$ & $\frac{\Delta m}{m} = - 3.4 \%$ &  & & & \\
 &$\frac{\Gamma_{\Phi}(\rho_0)}{\Gamma_{\Phi}}= 3.6 $ & & & & \\
\hline
\end{tabular}
}
\end{table}

\section{Conclusion}
The results on medium modifications of mesons
reported in the previous sections 
are summarized in table 1.
Dropping masses associated with a sizable broadening are found for the 
$\omega$ and $\Phi$ meson.For the $\rho$ meson, 
conflicting results are reported which need to be further clarified. 
Medium effects in nucleus-nucleus 
collisions at SPS energies rule out \cite{Damjanovic} or at least disfavour 
\cite{CERES_06} a universal downscaling of hadron masses in the nuclear medium. When comparing the experimental results it should be noted
that some detector systems have no or little acceptance for low meson momenta 
for which strong medium modifications are expected. 
The observed differences in the in-medium behaviour of $\rho$, 
$\omega$, and $\Phi$ mesons may be attributed to their different 
coupling to the nuclear
medium because of their isovector and isoscalar nature, respectively.

\section{Acknowledgements}
I would like to thank P. Braun-Munzinger, S. Damjanovic, C. Djalali, H. En'yo,
R. Muto, M. Naruki, R. Nasseripour, H.J. Specht, and  J. Stachel for 
discussions on their results. Discussions with S. Leupold, U. Mosel,
P. M\"uhlich and W. Weise on
the theoretical aspects of this work and with my colleagues
in the CBELSA/TAPS collaboration, in particular M. Kotulla and D. Trnka, are
highly appreciated. This work was supported by DFG through SFB/TR16
``subnuclear struture of matter''.

\end{document}